\documentstyle[11pt,emulateapj]{article}
\def\capt{\small \baselineskip 12pt }
\def\be{\begin{equation}}
\def\ee{\end{equation}}
\def\se#1{\S\ref{sec:#1}}
\def\fig#1{Figure~\ref{fig:#1}} 
\def\equ#1{Eq.~(\ref{eq:#1})}

\def\bm{{\bf m}}
\def\bd{{\bf d}}
\def\br{{\bf r}}
\def\etal{{\it et al. }}

\def\eg{{\it e.g.}}
\def\ie{{\it i.e.}}

\def\apriori{{\it a priori }}
\def\kms{\,{\rm km\,s^{-1}}}
\def\hmpc{\,{\rm h^{-1}Mpc}}
\def\hkpc{\,{\rm h^{-1}kpc}}
\def\ihmpc{\,(h^{-1} {\rm Mpc})^{-1}}
\def\kmsmpc{\,{\rm km\,s^{-1}Mpc^{-1}}}
\def\3hmpc{\,(h^{-1} {\rm Mpc})^3}
\def\ln{{\rm ln}}
\def\log{{\rm log}}

\def\la{\langle} \def\ra{\rangle}

\def\omm{\Omega_m}
\def\oml{\Omega_\Lambda}
\def\omb{\Omega_b}

\def\lcdm{$\Lambda$CDM}

\def\h60{h_{60}}
\def\P{{\cal P}}
\def\L{{\cal L}}
\def\kb{k_{\rm b}}
\def\sigv{\sigma_{\rm v}}

\begin{document}

\title{COSMOLOGICAL DENSITY AND POWER SPECTRUM FROM PECULIAR VELOCITIES: 
NONLINEAR CORRECTIONS AND PCA}

\author{L. Silberman$^1$, A. Dekel$^1$, A. Eldar$^1$ \& I. Zehavi$^{2}$}  

\affil{$^1$Racah Institute of Physics, The Hebrew University, Jerusalem 91904,
Israel}

\affil{$^2$NASA/Fermilab Astrophysics Group, Fermi National Accelerator 
Laboratory, Box 500, Batavia, IL 60510-0500, USA} 


\begin{abstract}

We allow for nonlinear effects in the likelihood analysis of galaxy peculiar 
velocities, and obtain $\sim 35\%$-lower values for the
cosmological density parameter $\omm$ and for the amplitude of mass-density 
fluctuations $\sigma_8 \omm^{0.6}$.
This result is obtained under the assumption that 
the power spectrum in the linear regime is 
of the flat \lcdm\ model ($h=0.65$, $n=1$, COBE normalized) 
with only $\omm$ as a free parameter.  
Since the likelihood is driven by the nonlinear regime, 
we ``break" the power spectrum at $\kb\sim 0.2\ihmpc$
and fit a power-law at $k>\kb$. This allows for independent 
matching of the nonlinear behavior and an unbiased fit in the linear regime.
The analysis assumes Gaussian 
fluctuations and errors, and a linear relation between velocity and density. 
Tests using mock catalogs that properly simulate nonlinear effects 
demonstrate that this procedure results in a reduced bias and a better fit.  
We find for the Mark III and SFI data 
$\omm=0.32 \pm 0.06$ and $0.37 \pm 0.09$ respectively,
with $\sigma_8 \omm^{0.6}=0.49 \pm 0.06$ and $0.63 \pm 0.08$,
in agreement with constraints from other data.
The quoted 90\% errors include distance errors and cosmic variance,
for fixed values of the other parameters. 
The improvement in the likelihood due to the nonlinear correction
is very significant for Mark III and moderately significant for SFI.

When allowing deviations from \lcdm, we find an indication for a 
wiggle in the power spectrum: an excess near $k\sim 0.05\ihmpc$
and a deficiency at $k \sim 0.1\ihmpc$ --- a ``cold flow".
This may be related to the wiggle seen in the power spectrum from 
redshift surveys and the second peak in the CMB anisotropy.


A $\chi^2$ test applied to modes of a Principal Component Analysis (PCA)
shows that the nonlinear procedure improves the goodness of fit 
and reduces a spatial gradient that was of concern in the purely linear 
analysis.  The PCA allows us to address spatial features of the data and to 
evaluate and fine-tune the theoretical and error models.  
It demonstrates in particular that the models used are appropriate for the 
cosmological parameter estimation performed.
We address the potential for optimal data compression using PCA.

\end{abstract}



\section{INTRODUCTION}
\label{sec:intro}

Our standard cosmological framework assumes that structure originated from
small-amplitude density fluctuations that were amplified by
gravitational instability.  These initial fluctuations are assumed to
have a Gaussian probability distribution, fully characterized by their
power spectrum $P(k)$.  On large scales, the fluctuations are expected
to be linear even at late times, still characterized by the initial
$P(k)$, which is directly related to the cosmological parameters. 
This makes the $P(k)$ a useful statistic for the study of both,
the origin of large-scale structure and the global cosmological parameters.

The $P(k)$ as estimated from galaxy redshift surveys 
(see reviews by Strauss \& Willick 1995; Strauss 1999)
is contaminated by unknown ``galaxy biasing", reflecting the possibility
that the spatial distribution of galaxies is not an accurate tracer
of the underlying {\it mass} distribution (\eg, recent references 
Blanton \etal 1999; 
Dekel \& Lahav 1999; Somerville \etal 2001; Tegmark \& Bromley 1999). 
Additional complications arise from redshift distortions,
triple-value zones and the nonlinearity of the density field,
which complicates the recovery of $P(k)$.
To avoid galaxy biasing, it is
advantageous to estimate the mass $P(k)$ directly from purely
dynamical data such as the peculiar velocities.  Another advantage of 
velocity over density data
is that they probe the density field on scales larger than the sample
itself, and therefore are subject to weaker nonlinear effects.

Direct estimation of the $P(k)$ from the reconstructed velocity fields 
by POTENT-like procedures (Dekel \etal 1999) 
is complicated by the need to correct for the effects 
of large noise, smoothing, and sparse and nonuniform sampling (\eg, Kolatt 
\& Dekel 1997; see also Park 2000).  
On the other hand, the likelihood analysis 
applied here, improving on the simplified linear version of
Zaroubi \etal (1997, Z97) and Freudling \etal (1999, F99),
acts on the `raw' data without pre-processing.
It utilizes much of the information content of the data, while
taking into account the measurement errors and the finite, discrete sampling.
The simplifying assumptions made 
in turn are that the peculiar velocities are drawn from a
{\it Gaussian\,} random field, that the velocity correlations can be derived
from the density $P(k)$ using {\it linear\,} theory, 
and that the errors 
are also {\it Gaussian}. The method requires to assume as a prior {\it model\,}
a parametric functional form for the $P(k)$, which then allows for
cosmological parameter estimation.

Since we address here the mass-density power spectrum as derived from
peculiar-velocity data, we determine directly the quantity $P(k)\,\omm^{1.2}$ 
(where $\omm$ is the cosmological mass-density parameter).  
This leads to a measure of a purely dynamical parameter such as
$\sigma_8 \omm ^{0.6}$ (where $\sigma_8$ is the
rms mass-density fluctuation in a top-hat sphere of radius
$8\hmpc$). When assuming a parametric functional form for the
mass $P(k)$, \eg, based on a cosmological CDM model, we could in principle
remove the degeneracy between $\omm$ and $\sigma_8$, and
determine a combination of dynamical parameters such as $\omm$,  
the baryonic contribution $\omb$,
the Hubble constant $h$, and the power index on large scales $n$
[where $P(k) \propto k^n$].  These parameters enter via the shape and amplitude 
of $P(k)$ as well as the geometry and dynamics of space-time.
 
Note for comparison that 
investigations involving galaxy redshift surveys commonly measure a
different parameter that does involve galaxy {\it biasing},
$\beta\equiv\Omega ^{0.6}/b$ (where $b$ is the linear biasing parameter). 
The parameters $\sigma_8 \omm ^{0.6}$ and $\beta$ (at $8\hmpc$) are related via
$\sigma_{8\rm g}$, referring to the rms fluctuation in the galaxy
number density.  Numerous measurements of $\beta$ have been carried
out so far, either based on redshift distortions, e.g., in IRAS catalogs
(Fisher \etal 1994; Tadros 1999; Hamilton, Tegmark \& Padmanabhan 2000)  
or based on comparisons of such  
redshift surveys and the peculiar-velocity data. Most recent
velocity-velocity comparisons found values for $\beta$ in the range
$0.4-0.7$  
(Davis, Nusser \& Willick 1996; Willick \etal 1997b; da
Costa \etal 1998; Kashlinsky 1998; Willick \& Strauss 1998;
Branchini \etal 2000),
while density-density comparisons have lead to values as high as 0.9
(e.g. Sigad \etal 1998).  A determination of the biasing-free quantity
$\sigma_8 \omm ^{0.6}$ directly from the peculiar velocity data may 
help clarifying the confusion about $\beta$. Moreover, the direct measure of
$\omm$ will enable a biasing-free result.

We use two catalogs of galaxy peculiar velocities.  
The Mark III (M3) catalog (Willick \etal 1995, 1996, 1997a) 
contains $\sim\!3000$ galaxies within 
$\sim\!70 \hmpc$.  It has been compiled from several different data sets 
of spiral and elliptical/S0 galaxies with
distances inferred by the forward Tully-Fisher (TF)
and $D_n\!-\!\sigma$ methods.
The sampling is dense nearby and much sparser at large distances.
The error per galaxy is on the order of $15-21\%$ of the distance.
The galaxies were first grouped into $\sim 1200$ objects, ranging from isolated
galaxies to rich clusters, in order to reduce the non-linear noise
and the resulting Malmquist bias. The data were then 
systematically corrected for Malmquist bias.
The SFI catalog (Haynes \etal 1999a, 1999b)
consists of $\sim\!1300$ late-type spiral galaxies with I-band
TF distances from two datasets.
It covers a volume similar to M3, with sparser sampling nearby
but a more uniform coverage of the volume.
Following da Costa \etal (1996), about $7\%$ of the galaxies, those
with the smallest line-width ($\log w \le 2.25$), have been discarded 
because of the
unreliability of the TF relation and its scatter at such line-widths. 
The data were corrected for Malmquist bias using the method described
in Freudling \etal (1999).

In earlier papers (Z97; F99; Zehavi \& Dekel 1999), 
we have applied to these data a purely linear likelihood analysis.
The model assumed was a {\it linear\,} $P(k)$ on all scales,
taken at large from the family of CDM models, normalized
by COBE's measurements of the large-scale fluctuations in the CMB.
The free parameters were typically $\omm$, $n$, and $h_{65}$ 
(the Hubble constant in units of $65\kmsmpc$).
The constraints obtained from the two data sets turned out to be similar,
both yielding a relatively high $P(k)$, in general agreement
with the direct estimate from the ``POTENT'' reconstruction (Kolatt
\& Dekel 1997). The constraints
defined an elongated two-dimensional surface in the $\omm$-$h$-$n$ parameter
space, which could be crudely approximated in the case of a flat universe
(and no tensor fluctuations), for M3 and SFI respectively, 
by $\omm\, h_{65}^{1.3}\, n^2 \simeq 0.56 \pm 0.09$ and $0.51 \pm 0.10$ 
($90\%$ errors). 
Corresponding constraints were  
$\sigma_8 \omm^{0.6} \simeq 0.85 \pm 0.11$ and $0.82 \pm 0.12$ respectively.
These results seemed to be conservatively consistent with the $2\sigma$ 
lower bounds of $\omm > 0.3$
obtained from peculiar velocities by other biasing-free methods 
(Nusser \& Dekel 1993; Dekel \& Rees 1994; Bernardeau \etal 1995),  
but they imply higher values for $\omm$ and $\sigma_8$
than obtained from other estimators, e.g. based on cluster abundance
($\sigma_8\omm^{0.56} \simeq 0.57\pm 0.05$, White, Efstathiou \& Frenk 1993)
or its evolution
($\omm \simeq 0.45\pm 0.25$ and $\sigma_8 \simeq 0.7\pm 0.15$, Eke \etal 1998).

The likelihood method has been tested by Z97 and F99 
using mock catalogs drawn from an N-body simulation 
of a constrained realization of our real cosmological neighborhood
(Kolatt \etal 1996). These tests indicated that nonlinear effects may cause 
only small differences in the results. 
However, this simulation
was limited in an important way; it
had a fixed dynamical resolution of only $\sim 2 \hmpc$, and therefore
suffered from certain smearing of nonlinear effects on the scales of 
individual galaxies and close pairs.  Being a simulation of an 
$\omm=1$ cosmology
also contributed to the underestimation of the density fluctuations
associated with the observed peculiar velocities, compared to
the currently favored cosmology with a lower $\omm$.
Despite the fact that the nonlinearities are 
expected to be weaker for velocities than for densities, and that the flows are
known to be relatively ``cold" on small, mildly nonlinear scales, 
it is quite possible
that the resolution of the early simulation was insufficient for an accurate 
evaluation of how nonlinear effects influence the results in the real world.
 
We have therefore generated new mock catalogs that are based on simulations 
of much higher resolution (Kauffmann \etal 1999a).
We simulated both a high-$\omm$ model and a low-$\omm$ one, and
galaxies were identified based on a more physical semi-analytic
scheme (Kauffmann \etal 1999a, 1999b),  which allows us to better mimic 
the real sampling and correct for associated biases.

Equipped with these nonlinear mock catalogs, we re-consider the nonlinear
effects in our original linear analysis.  Once we discover an indication for
a bias in this analysis, we introduce ways to incorporate nonlinear effects.  
We realize that the fit is driven by the small scales, because close pairs 
arise from nearby galaxies of small errors.  This 
means that even weak non-linear effects on small scales may bias the results.  
We do not have yet a good analytic approximation for the nonlinear 
corrections to the velocity $P(k)$, so we simply add free parameters 
that allow independent matching of the nonlinear behavior. For example, 
we introduce a {\it break\,} in $P(k)$ at $\kb \sim 0.2$, allow an arbitrary 
two-parameter power-law fit in the nonlinear regime $k>\kb$ and thus free the
linear part of the spectrum at $k<\kb$, and the associated cosmological
parameters, to be determined unbiased.
 
This nonlinear correction procedure is first tested using the nonlinear
mock catalogs, and then applied to the M3 and SFI data. We find that the 
obtained values of $\omm$ and $\sigma_8$ are significantly lower 
than in the purely 
linear analysis, and the results are not sensitive to the exact way by 
which the nonlinear effects are incorporated.

We also investigate the power spectrum in the relevant range of scales
independent of a specific cosmological model, 
by allowing as free parameters the actual values of $P(k)$
in finite intervals of $k$ (also in Zehavi \& Knox, in preparation). 
In particular, this allows a
detection of marginally significant deviations from the 
\lcdm\ power spectrum, which can be characterized as ``cold flows".

The likelihood analysis provides only relative likelihood of the 
different models, not an absolute goodness of fit (GOF).
An indication for a problem in the goodness of fit in the purely linear 
analysis came from a $\chi^2$ estimate in modes of a 
principal component analysis 
(Hoffman \& Zaroubi 2000). It seems to be associated with a
problem noticed earlier by Freudling \etal (1999), of a spatial gradient 
in the obtained value of $\omm$.
We develop a method based on $\chi^2$ and PCA 
as a tool for evaluating the goodness of fit in our procedure, 
and find a significant improvement in the GOF when the nonlinear 
corrections are incorporated and the most noisy data are pruned.

In \se{method} we describe the likelihood
method of analysis, the parametric models used as priors,
and the way we allow for nonlinear effects.
In \se{test} we test and calibrate the method using mock catalogs.
In \se{results} we present the resultant power spectrum and 
the constraints on the cosmological parameters for \lcdm, and detect
hints for deviations from this model. 
In \se{pca} we address the goodness of fit via $\chi^2$ in modes of PCA.
In \se{conc} we conclude.

\section{METHOD}
\label{sec:method}

\subsection{Likelihood Analysis}
\label{sec:method_like}

The general method has been developed and applied in Zaroubi \etal (1997)
and Freudling \etal (1999) (following Kaiser 1988; Jaffe \& Kaiser 1994).
The goal is to estimate the power spectrum of mass density
fluctuations from peculiar velocities, by finding maximum likelihood
values for parameters of an assumed model power spectrum.
Given a data set $\bd$, our objective is to estimate the most likely
model parameters $\bm$.
Using Bayes theorem the conditional probabilities are related by
\be
\P (\bm \vert \bd ) = {\P(\bm) \P(\bd|\bm) \over \P(\bd)} \,,  
\ee
and assuming a uniform prior $\P(\bm)$,
the task becomes the maximization of the the likelihood function 
$\L=\P(\bd|\bm)$
as a function of the assumed model parameters.

Under the assumption that both the underlying velocities and the
observational errors are independent Gaussian random fields,\footnote{
The assumed Gaussianity of the velocity field in the mildly nonlinear
regime is supported by simulations (Kofman \etal 1994, Kudlicki \etal 2001),
and is verified by the nearly normal distribution of the observed $\ln(z/d)$
in our sample.} 
the likelihood function can be written as
\be
\label{eq:like}
\L = {1 \over [ (2\pi)^n \det(C)]^{1/2} }
  \exp\left( -{1\over 2}\sum_{i,j}^n {u_i C_{ij}^{-1} u_j}\right)\,.
\ee
This is simply the corresponding multivariate Gaussian distribution,
where $\{u_i\}_{i=1}^{n}$ is the data set of $n$ observed peculiar
velocities at locations $\{\br_i\}$, and $C$ is their correlation
matrix. Expressing each data point as the sum of the actual signal and
the observational error $u_i=s_i+n_i$, the elements in the
correlation matrix have two contributions:
\be
C_{ij}\equiv \la u_i u_j \ra = \la s_i s_j \ra + \la n_i n_j \ra
\equiv S_{ij}+ N_{ij} \ .
\label{eq:C}
\ee
The first term is the correlation matrix of the signal,
which is calculated from the theoretical $P(k)$ model at the sample 
positions $\br_i$.  The second term is the error matrix, which is 
diagonal based on the assumption that the distance errors of the
objects in the sample are uncorrelated
with each other.  This should be true for the two components of the errors,
the observational errors and the intrinsic scatter of the TF
relation.\footnote{ 
Freudling \etal (1999) tested the impact of uncertainties in the bias 
correction, which might have introduced correlations in the  errors, 
by varying parameters in the bias model within the expected uncertainties. 
They found the changes in the results to be negligible compared to the other 
systematic random errors in the analysis.}  

For a given $P(k)$, the signal terms are
calculated using their relation to the parallel and perpendicular
velocity correlation functions, $\Psi_{\Vert}$ and $\Psi_{\perp}$,
\be
S_{ij}=\Psi_{\perp}(r)\sin\theta_i \sin\theta_j +
\Psi_{\Vert}(r)\cos\theta_i
\cos\theta_j  \, ,
\ee
where $r=\vert \br \vert=\vert \br_j-\br_i \vert$ and the angles are
defined by $\theta_i=\hat{\br_i}\cdot\hat{\br}$ (G\'orski 1988; Groth,
Juszkiewicz \& Ostriker 1989).  In linear theory, each of these can be
calculated from the $P(k)$,
\be
\Psi_{\perp,\Vert}(r)= {H_0^2 f^2(\omm )\over 2 \pi^2}
\int_0^\infty P(k)\, K_{\perp,\Vert}(kr)\, dk \,,
\ee
where $K_{\perp}(x) = j_1(x)/ x$ and $K_{\Vert}(x) = j_0-2{j_1(x)/
x}$, with $j_l(x)$ the spherical Bessel function of order $l$.  The
cosmological $\omm$ dependence enters as usual in linear theory via
$f(\omm)\simeq \omm^{0.6}$, and $H_0$ is the Hubble constant 
($H_0 \equiv 100 h \kmsmpc$).

For each choice of the model parameters
the correlation matrix $C$ is computed, inverted, and substituted in the
likelihood function [\equ{like}]. Exploring the chosen
parameter space, we find the parameters for which the likelihood is
maximized.\footnote{
Note that since the model parameters appear also in the
normalizing factor of the likelihood function, through $C$, maximizing
the likelihood is {\it not\,} equivalent to minimizing the $\chi^2$.}
The main computational effort is the calculation and inversion of the
correlation matrix $C$ in each evaluation of the likelihood. It is an
$n \times n$ matrix, where the number of data points $n$ is typically
more than $1000$. This number is expected to increase when future samples
become available, which will require a procedure for data compression
(see \se{pca}).

The random measurement errors deserve special attention; they
add in quadrature to the true $P(k)$ and thus
propagate into a systematic uncertainty in the results.
Zaroubi \etal (1997) used \apriori\ estimates of the errors,
which were
based on evaluations of the observational and internal scatter of the TF
distances using galaxies in clusters or local velocity-field models (Willick
\etal 1995).
Freudling \etal (1999) improved the method by incorporating
the errors into the likelihood
analysis itself via an error model with free parameters,
which only weakly builds upon the original error estimates.
The maximum-likelihood errors were found to be within 5\% of the
\apriori\ error estimates, thus allowing us to adopt the \apriori\ error
estimates in our following analysis.

Relative confidence levels are estimated by approximating $-2\ln\L$ as a
$\chi^2$ distribution with respect to the model parameters.  The
likelihood analysis provides only relative likelihoods of different
models. Absolute goodness of fit is addressed separately in \se{pca} below.

\subsection{The Cosmological Power Spectrum Model}
\label{sec:method_models}

In the linear regime, we use as prior the parametric form for the $P(k)$ 
based on the general CDM model,
\be
\label{eq:cdm}
P(k) = A_{\rm c}(\omm,\oml,n)\
T^2(\omm,\omb,h; k)\ k^n\,,
\ee
where $A_{\rm c}$ is the normalization factor and
$T(k)$ is the CDM transfer function proposed by Sugiyama (1995,
a slight modification of Bardeen \etal 1986):

\begin{eqnarray}
\label{eq:Tcdm}
\lefteqn{ T(k) =  {\ln\left(1+2.34q \right ) \over 2.34q} \times } \nonumber \\
& &\left[1+3.89q+(16.1q)^2+(5.46q)^3+(6.71q)^4\right]^{-1/4} \,, 
\nonumber \\ & &
\end{eqnarray}

\be
q=k\, [ \omm h \, \exp (-\omb - \sqrt{2h} \omb/\omm)] \ihmpc ]^{-1}\,.
\ee

We restrict ourselves in the present paper to the
flat cosmological model with a cosmological constant ($\omm+\oml=1$),
a scale-invariant power spectrum on large scales ($n=1$),
and no tensor fluctuations. The Hubble constant is fixed at $h=0.65$ 
(\eg, based on Freedman 1997).

As in earlier power-spectrum analyses,
the baryonic density is set to be $\omb=0.024 h^{-2}$ (Tytler, Fan \&
Burles 1996), and the amplitude $A_{\rm c}$ is fixed by the COBE 4-year data 
(Hinshaw \etal 1996; G\'orski \etal 1998),
\begin{eqnarray}
\label{eqn: COBE-norm}
\lefteqn{\log A_{\rm c}=7.84 -8.33\omm +21.31\omm ^{2}-29.67\omm ^{3}+}\nonumber \\
& & 10.65\omm ^{4} +15.42\omm ^{5}- 6.04\omm ^{6}-13.97\omm ^{7}+ \nonumber \\
& & 8.61\omm ^{8}  \, .
\end{eqnarray}

In order to check the sensitivity to uncertainties in these quantities,
we have repeated the likelihood analysis with a later estimate of
$\omb=0.019 h^{-2}$ (Burles \etal 1999) and an alternative COBE
normalization (Bunn \& White 1997, equations 25 and 29). 
The obtained values of $\omm$ are found to be very robust to these changes, 
with variations smaller than 2\%.

\subsection{Broken Power Spectrum}
\label{sec:method_break}

As will be demonstrated below (\se{test}),
the maximum-likelihood solution is driven by the  
small scales, $k>0.2\ihmpc$, because close pairs preferentially
consist of nearby galaxies for which the errors are typically small.
In the case where the same parameters determine the $P(k)$ on all scales,
this means that even small inaccuracies in the power spectrum shape
at large wave-numbers may bias the results at small wave-numbers, 
and therefore the value of $\omm$.  
An accurate nonlinear correction to the linear 
{\it velocity\,} power spectrum could have been very useful
in avoiding this bias, but, unfortunately, such a correction
is not yet available. 
As mentioned earlier,
a successful empirical approximation does exist for the nonlinear 
correction to the {\it density\,} $P(k)$ (Peacock \& Dodds 1996, PD), 
modeling a gradual deviation from the linear $P(k)$ at $k>0.2\ihmpc$,
but the generalization to a velocity correction is not straightforward
because there is no explicit exact relation between velocity and density
in the nonlinear regime.\footnote{First attempts in this
direction are made by Sheth, Zehavi \& Diaferio (2000).} 

The procedure adopted here is to detach the nonlinear regime
from the linear regime by introducing a ``break" in the power spectrum at
a wavenumber $\kb$. We then assume the \lcdm\ shape for the $P(k)$
at $k<\kb$, 
determined by physical free parameters such as $\omm$, and allow an almost 
arbitrary function with enough free parameters to fit the data at $k>\kb$. 
We try, for example, a power law, with two free parameters: $P(k)=Bk^{-s}$.
This power-law serves the purpose of feeding the ``likelihood monster"
residing in the nonlinear regime, while freeing the linear part of the 
spectrum, and the associated cosmological parameters, to be determined 
unbiased.
The break scale could be an additional free parameter;
we test below the robustness of the results to the actual choice
of $\kb$.

This approach can be carried to an extreme where we break $P(k)$
in several places, and fit arbitrary functional forms independently
within finite intervals of $k$. Once we do that, we allow for more
flexibility in the power-spectrum shape, and can detect
specific deviations from the predicted CDM shape. Naturally, this procedure 
would limit our ability to address cosmological parameters.
The choice of a series of independent step functions 
(or ``band powers", forming a histogram) is especially appealing 
computationally,
because in this case the correlation matrix becomes a simple linear 
combination of the correlation matrices of the individual segments, and then
the integrals entering the correlation matrix need to be computed only
once.\footnote{
A way to implement band powers with a large
number of free parameters is via an iterative quadratic
estimator scheme, commonly applied to CMB measurements (\eg, Bond, Jaffe
\& Knox 1998), which greatly improves the computational efficiency and 
simultaneously provides the cross correlation between the different bands 
(Zehavi \& Knox in preparation).}

Alternatively, we can repeat the procedure of Freudling \etal (1999)
(also used in Bridle \etal 2001) 
where the nonlinear effects are accounted for by adding to the linear
velocity correlation model a free parameter of uncorrelated velocity 
dispersion at zero lag, $\sigv$, e.g., representing small-scale 
random virial motions.
This is a different way of incorporating nonlinear effects, possibly with 
a different physical interpretation.  It would thus be interesting to see
how the different methods affect the results in the linear regime,
and whether both are needed.
%

It is not clear that such prescriptions will capture the exact shape of the 
power spectrum in the nonlinear regime, and this is not our purpose here. 
However, we find that they may be sufficient for more accurate
estimation of the power spectrum in the linear regime, 
and for an unbiased determination of the cosmological density parameter
under an assumed parametric shape for the linear power spectrum.
%

\section{TESTING THE METHOD}
\label{sec:test}

\subsection{Mock catalogs}
\label{sec:test_mock}

We test the method using artificial mock catalogs based on a 
cosmological simulation, in which the ``true" cosmological parameters and 
linear power spectrum are fully known \apriori\ and where nonlinear effects  
are simulated with adequate accuracy on galactic scales.
We use the unconstrained ``GIF" simulation (Kauffmann \etal 1999a)
of the flat \lcdm\ cosmology with $\omm=0.3$.  
The initial fluctuations in this simulation
were Gaussian, adiabatic and scale-invariant, 
$n=1$. The $P(k)$ shape parameter was $\Gamma=\omm h=0.21$ 
(namely $h=0.7$) and the amplitude is such that $\sigma_8=0.9$ 
(extrapolated by linear theory from the initial conditions to $z=0$), 
consistent with both the present
cluster abundance and COBE's measurements on large scales.
The $N$-body code is a version of the adaptive particle-particle
particle-mesh (AP$^3$M) Hydra code developed as part of the VIRGO
supercomputing project (Jenkins \etal 1998). 
The simulation has $256^3$
particles and $512^3$ cells, and a gravitational softening
length of $30 \hkpc$, inside a box of side $141.3\hmpc$.

Dark-matter halos were identified using
a friends-of-friends algorithm with a linking length of 0.2 (corresponding
to a density contrast of $\sim 125$ at the halo edges)
and a minimum of 10 particles per halo was imposed.
Luminous galaxies were planted in the halos based on a semi-analytic 
scheme (Kauffmann \etal 1999a, 1999b)  
whose main elements can be summarized as follows.
A merger history is constructed for each halo. The gas in every 
progenitor halo is assumed to cool radiatively and settle into a galactic disk.
Stars are assumed to form in a rate proportional to the mass of cold 
gas and inversely proportional to the dynamical time.
Cold gas may be re-heated by supernovae feedback and removed from the
disk or from the halo all together.
When halos merge, the central galaxy of the
largest halo becomes the central galaxy of the new halo and
all other galaxies become satellites which later
merge with the central galaxy on a dynamical-friction time scale. 
Major mergers result in destruction of disks and formation of
spheroids, thus determining the morphological type.
The star formation history of each galaxy is convolved with stellar
population synthesis models and extinction models to obtain total 
luminosities in different bands.

We assigned to each galaxy a linewidth based on the TF relation and
scatter assumed in Kolatt \etal (1996), and a diameter based on the
magnitude-diameter relation used in Freudling \etal (1995).
We then generated 10 mock catalogs which resemble the M3 
catalog and 10 which resemble the SFI 
sample. The selection procedure was simulated
using the galaxy magnitudes (M3) or angular diameters (SFI) generated above,
in a way that reproduces the redshift and luminosity distributions
in the real catalogs. The samples were further randomly diluted, simulating
selection by other independent properties such as inclination, to
match the number of galaxies in the real catalogs. This random
sampling, along with the random distance errors (introduced by the TF
scatter), has been repeated 10 times to generate the 10 mock catalogs. 
The mock data were corrected for Malmquist biases
exactly the way they were corrected in the real data, including
the first step of grouping for the M3 samples.

The degree of {\it nonlinearity\,}, 
which is a key feature in our current analysis,
depends on the exclusion of clustered galaxies from the sample.
In M3, rich clusters of either elliptical or spiral galaxies
were selected \apriori\ and considered as single massive objects moving
with their center of mass velocities. The typical radii of rich clusters
are crudely $\sim 3\hmpc$ and $\sim 6 \hmpc$ for ellipticals and spirals 
respectively.
Since the M3 catalog is densely sampled at small distances, and it
includes elliptical galaxies which tend to cluster,
further groups were identified from the ``field" samples and were
also treated as single objects.
In the SFI catalog, which consists only of spirals, the cluster galaxies were 
excluded \apriori\ and included in an associated catalog, termed SCI; 
there was no need for further grouping of the
relatively sparse field sample.
Unfortunately, the exclusion and grouping of clustered galaxies
in the two real catalogs
did not follow a simple uniform and objective algorithm that is
straightforward to mimic in the simulations.
 
\def\rc{r_{\rm c}}
We therefore produced a suite of 10 sets of 10 mock catalogs each,
spanning a range of degree
of nonlinearity, created by varying the criterion for the exclusion of
cluster galaxies. Galaxies were excluded if they lie within a distance
$\rc$ from the cluster center. The ``linearity parameter" $\rc$ is
measured in units of $3.5\hmpc$
and $1.5\hmpc$ for spirals and ellipticals respectively (the radii
used in the old mock catalogs of Kolatt \etal 1996), and it ranges from
$\rc=0.1$ to $1$ in steps of $0.1$.
This allows us to study how our method performs in the presence
of different degrees of nonlinear effects.

\subsection{Bias in $\omm$ and its Correction}
\label{sec:test_test}

\begin{figure*}[t!]
\vspace{10.5truecm}
\includegraphics{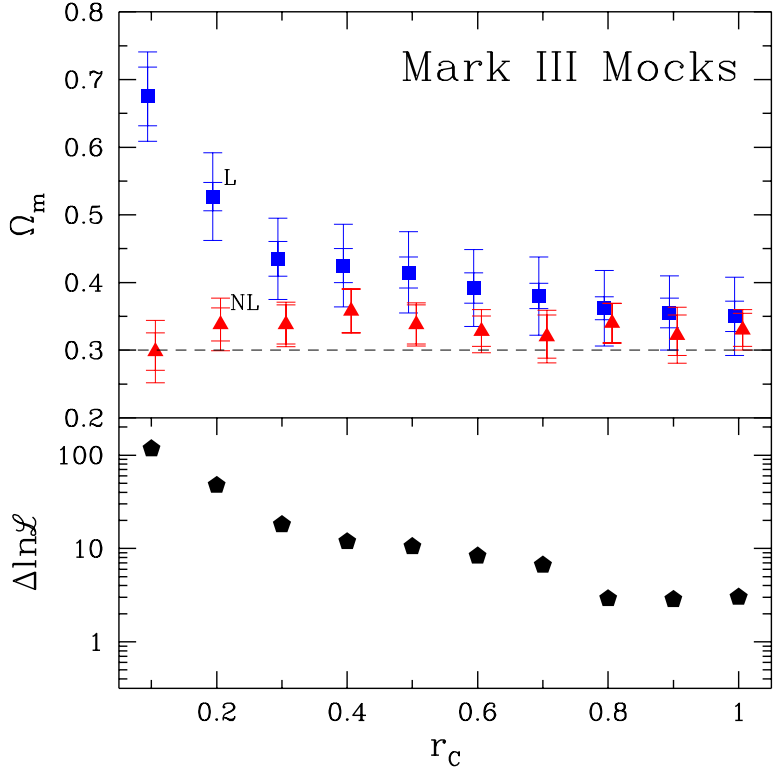}
\caption{\protect\capt
Testing the method at different levels of nonlinearity.
Bottom: the value of the recovered density parameter $\omm$
as a function of the degree of linearity of the dataset, as measured by
$\rc$.
The ``true" value is $\omm =0.3$.
Each symbol marks the average over 10 mock M3 catalogs,
and the inner error-bar marks the corresponding standard deviation.
The outer error-bar is the $84\%$
likelihood uncertainty (corresponding to $\Delta\ln\L=1$),
which includes cosmic variance.
The squares represent the results of the likelihood analysis using
the purely linear \lcdm\ $P(k)$; they show a significant bias that
increases with decreasing $\rc$.
The triangles represent the results of the improved analysis using
a broken-\lcdm\ $P(k)$; the bias is drastically reduced.
Top: the corresponding mean improvement in $\log \L$
for the nonlinear analysis versus the linear analysis.
}
\label{fig:mock_rc}
\end{figure*}

\begin{figure*}[t!]
\vspace{11.2truecm}
\includegraphics{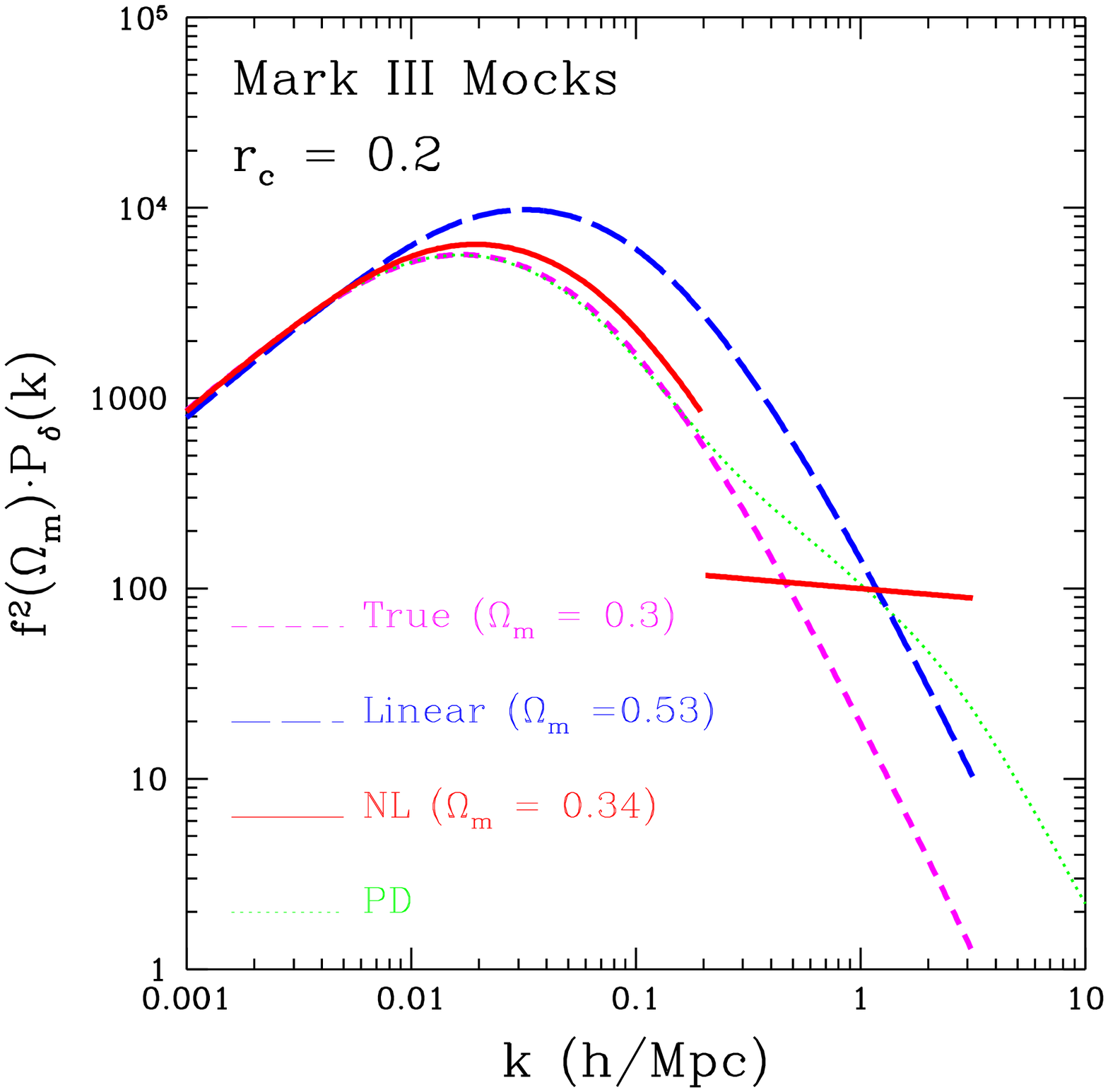}
\caption{\protect \capt
Mean power spectra recovered from the M3 mock catalogs of $\rc =0.2$.
The target is the true linear $P(k)$ (marked ``True").
The nonlinear correction by Peacock \& Dodds (PD) is shown for
comparison.
The result from the linear analysis is marked ``L".
The result from the nonlinear analysis with $\kb=0.2\ihmpc$ ,
marked ``NL", consists of a \lcdm\ function at $k<\kb$ and a power-law
at
$k>\kb$.
The $P(k)$ is in units of $\!\3hmpc$.
}
\label{fig:mock_pk}
\end{figure*}

We have applied the likelihood analysis to each of the $10 \times 10$ mock
M3 catalogs. The recovered values of $\omm$ are shown in \fig{mock_rc}
as a function of the degree of linearity of the dataset, as measured by $\rc$.
The ``true" target value is $\omm =0.3$.
Each symbol represents the average over the 10 mock catalogs for each value of
$\rc$. The small errorbars mark the standard deviation over these $10$ mock
catalogs, and thus represent the uncertainty due to
the random sampling and the random distance errors.
The large errorbars are 
$84\%$ errors based on the likelihood analysis,
namely determined by $\Delta \ln \L = 1$; they thus include 
the effects of random distance errors and cosmic variance.

We first apply the purely linear analysis, with the linear \lcdm\ power
spectrum at all scales, COBE normalized,
and with $\omm$ as the only free parameter 
while all the other parameters are fixed at their ``true" values.
We see that the linear likelihood analysis systematically overestimates the
value of $\omm$. As the data become more nonlinear, the recovered
value of $\omm$ becomes higher, and the bias more significant.
For example, at $\rc=0.2$, we obtain $\omm=0.53$, which is more than a 
$4\sigma$ deviation from the true value.

Next, we apply the improved procedure, allowing for a break in the $P(k)$
at $\kb=0.2\ihmpc$ and two additional free parameters in the nonlinear
regime. As demonstrated in \fig{mock_rc}, 
the bias is practically removed for all levels of nonlinearity.
The figure also shows the corresponding improvement in $\log \L$ 
when the linear analysis is replaced with the nonlinear analysis. 
The improvement grows continuously with decreasing $\rc$, 
from $\Delta \ln \L \sim 1 $ at $\rc=1$ to $\sim 120$ at
$\rc=0.1$.

\fig{mock_pk} shows the average mass-density  
power spectra recovered from the M3 mock
catalogs of linearity $\rc =0.2$. The true linear density $P(k)$ of 
the simulation, moved forward in time by linear theory,
is shown for comparison. Also shown is the Peacock \& Dodds (1996) 
nonlinear correction at large $k$.
The $P(k)$ recovered by the linear likelihood analysis (L), corresponding to
$\omm=0.53$, is higher than the true $P(k)$ at $k=0.2\ihmpc$
by a factor of $\sim 5$. The nonlinear analysis (NL), with $\omm = 0.34$
compared to the true $\omm=0.30$, brings the $P(k)$ down much closer
to the true $P(k)$.

The power-law segment at $k>\kb$ crudely recovers the amplitude 
of the PD power spectrum at $k \sim 1$, but apparently not the general slope.
A good agreement between the two is not obvious \apriori\ because
the PD correction refers to the density $P(k)$, while 
our likelihood analysis is based on the velocity power spectrum and
is still making use of the linear relation between velocity and density.
We shall see below that when applied to the real data, of either M3 or SFI,
the NL segment does match the PD approximation somewhat better.

The true initial power spectrum in the simulation was actually
based on the $\Gamma$ functional form (\eg, Efstathiou, Bond \& White 1992),
which is a slightly different approximation to the \lcdm\ spectrum than the one
used as a prior in our likelihood analysis, \equ{Tcdm}.
The differences between these two power spectra for the same
values of $\omm h$ are relatively small, \eg, at the level of 20\% at $k=0.1$.
To test the robustness of our results to these small differences in the
power-spectrum shape, we also applied the same
likelihood analysis to the mock catalogs using the $\Gamma$
model as prior. The free parameter in this case was 
$\Gamma$,  which is
equivalent to $\omm$ for a fixed $h$. The normalization of $f(\omm) P(k)$
was fixed at a small wavenumber, $k=0.001(\hmpc)^{-1}$, to equal the true
normalization in the simulation.
The results are found to be robust. For example, at a linearity 
of $\rc=0.2$, the linear $\Gamma$ model yields a best fit of $\omm =0.54$
(instead of $0.53$ when \equ{Tcdm} is used, with COBE normalization), 
and the broken $\Gamma$ model yields $\omm =0.36$ (instead of $0.34$).

When nonlinear effects are in action, one might worry about mode-mode
coupling affecting the results in the linear regime. 
The unbiased estimates of the linear $P(k)$ and $\omm$
obtained with our method when applied to the mock catalogs indicates that
the modes for $k<\kb$ are practically decoupled from modes with $k>\kb$.
We can assume that the linear modes with $k<\kb$ are
not coupled among themselves, but we do not know much about
possible coupling among the nonlinear modes with $k>\kb$.
This is fine as long as we do not assign physical significance to the 
shape of the power spectrum recovered on these scales,
beyond assuming that the overall probability distribution is well
approximated by a model with a certain function playing the role of the power
spectrum.
%

Our conclusion from the above test using the mock catalogs is that, 
in the presence of significant nonlinear effects in the data, 
the purely linear likelihood analysis might yield  
a biased estimate of $P(k)$ and $\omm$.
The broken-$P(k)$ analysis successfully eliminates the dependence 
of the results on the nonlinear effects and practically
corrects the bias in the results.

\section{RESULTS}
\label{sec:results}

\subsection{Broken \lcdm: the Value of $\omm$}
\label{sec:results_lcdm}

We now apply the improved likelihood analysis to the real data of M3 and SFI.
Similar to the tests with the mock data,
our \lcdm\ model is restricted to a flat universe with $h=0.65$,
$\omb h^2 =0.02$ and $n=1$, leaving only one cosmological parameter 
free to be determined by the maximum likelihood analysis, namely $\omm$.
Note that contrary to the situation in the mock catalogs we now do not know 
\apriori\ that the \lcdm\ model is the right one or  
that the values of the fixed parameters are the accurate ones. 
The purely linear analysis yields $\omm=0.56\pm 0.04 $ and $\omm=0.51\pm 0.05$ 
for M3 and SFI respectively ($90\%$ errors), consistent with Z97 and F99 
when $h$ and $n$ are fixed at the values quoted above.
Based on the test using mock catalogs,
we now suspect that these might be overestimates.

\begin{figure*}[t!]
\vspace{9.0truecm}
\includegraphics{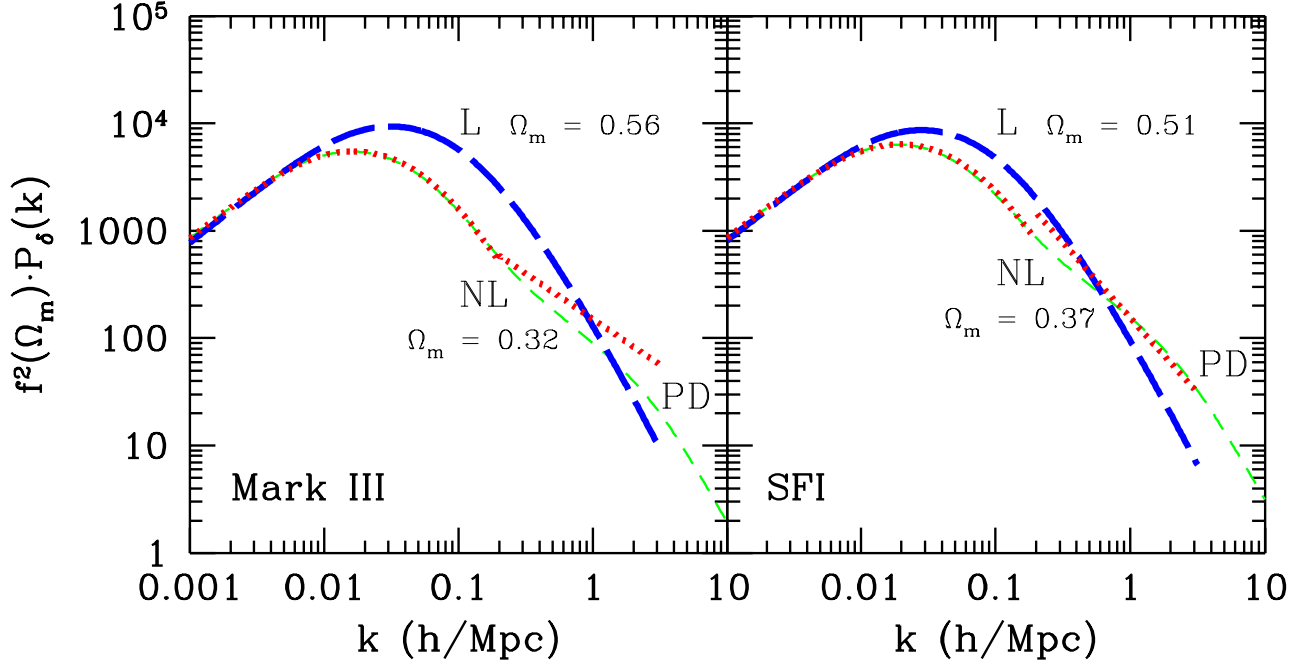}
\caption{\protect \capt
The recovered power spectra from the real data of M3 (left) and SFI (right). 
The $P(k)$ yielded by the purely linear analysis is marked ``L",
while the nonlinear analysis, with a break at $k=0.2\ihmpc$,
is marked ``NL".
Also shown for comparison is an extrapolation of the linear part
of the recovered $P(k)$ into the nonlinear regime by the PD approximation.
The $P(k)$ is in units of $\!\3hmpc$.
}
\label{fig:m3sfi}
\end{figure*}

\begin{figure*}[t!]
\vspace{6.0truecm}
\includegraphics{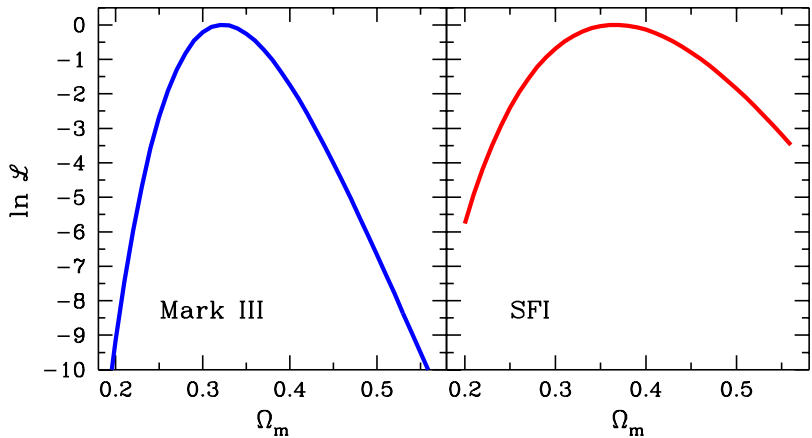}
\caption{\protect \capt 
Likelihood function for the values of $\omm$ due to the nonlinear
analysis with $\kb = 0.2\ihmpc$, from the real data of M3 (left)
and SFI (right).
}
\label{fig:likeomega}
\end{figure*}

\begin{figure*}[t!]
\vspace{11.0truecm}
\includegraphics{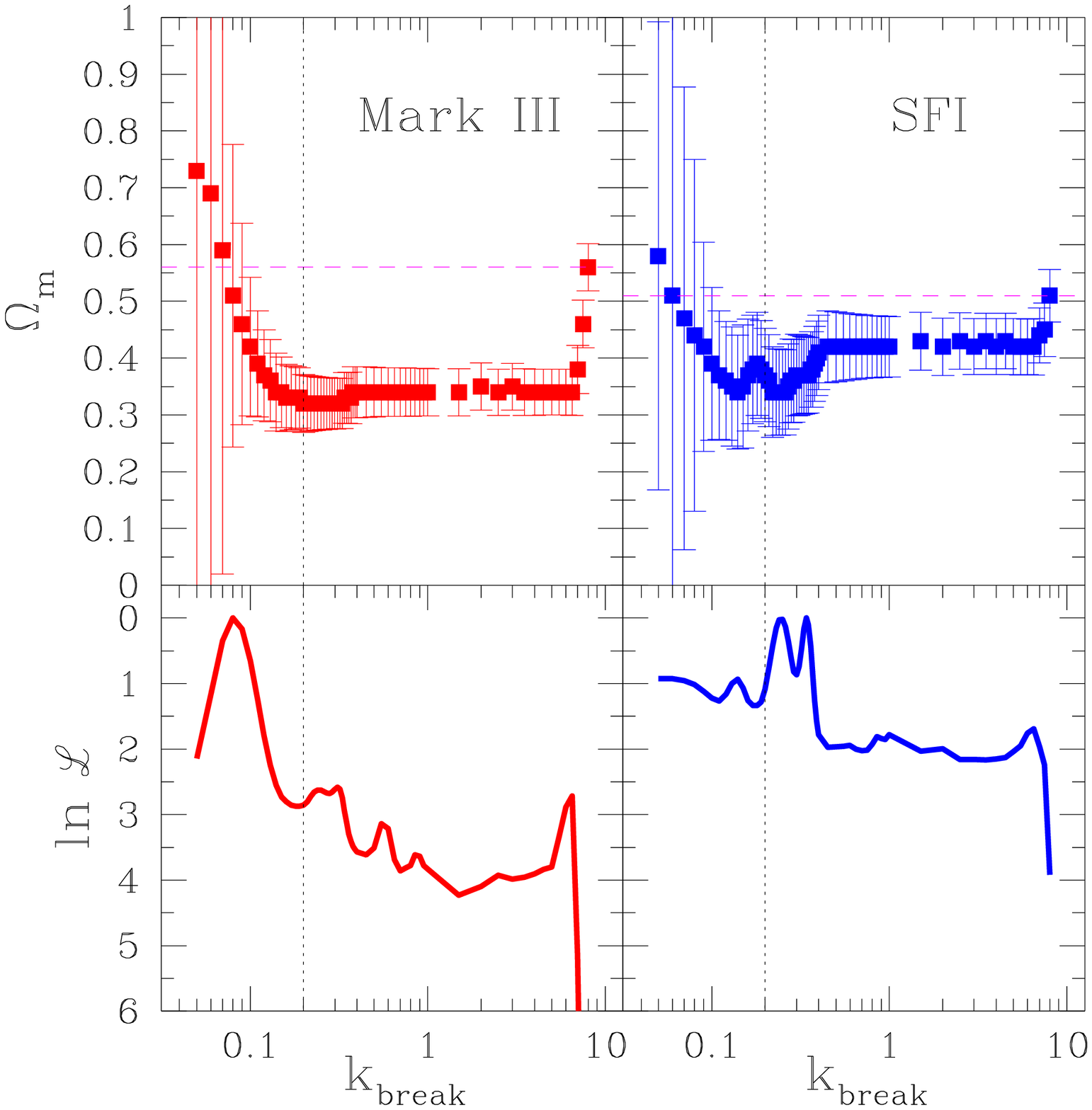}
\caption{\protect \capt
Robustness to the break scale.
The results of the nonlinear analysis with the broken-\lcdm power
spectrum,
for the real data, as a function of the break length $\kb$ in units of
$\!\ihmpc$.  The errors are $84\%$ from the likelihood function.
Bottom: the recovered value of $\omm$.
Top: the corresponding likelihood.
Left: M3. Right: SFI.
}
\label{fig:omega_by_k}
\end{figure*}

\fig{m3sfi} shows the maximum-likelihood power spectra 
as derived from the two catalogs of real data.
The linear analysis yields a high amplitude, corresponding
to a high value of $k_{\rm peak}$ where $P(k)$ is at maximum,
and the corresponding high value of $\omm$.
 
The nonlinear analysis with a break at $\kb=0.2\ihmpc$
yields a shift of $k_{\rm peak}$ towards lower $k$ values,
associated with a lower value of $\omm$,
and a corresponding lower amplitude for $P(k)$ in much of the linear regime.
The new values are $\omm=0.32\pm 0.06$ for M3 and $\omm=0.37\pm 0.09$ for SFI.
 
The corresponding best-fit values of
$\sigma_8\omm ^{0.6}$ are $0.49 \pm 0.06$ and $0.63 \pm 0.08$
for M3 and SFI respectively.  
These values are consistent with the estimates from cluster abundance
(\eg, Eke \etal 1998).

The change caused in the value of $\omm$ due to the nonlinear correction
is similar to the corresponding change in the mock catalogs of M3
at a relatively high degree of nonlinearity, $\rc \simeq 0.2$ in \fig{mock_rc}.

The best-fit power-law segments in the nonlinear regime
are $145 k^{-0.8}$ for M3 and $155 k^{-1.4}$ for SFI
(where $k$ is in units of $\!\ihmpc$, and $P(k)$ in units of $\!\3hmpc$.
The power-law segments roughly coincide
with the linear \lcdm\ segments at $\kb$, indicating that this broken
power spectrum is a sensible approximation to the actual shape of
the $P(k)$. 
Interestingly, the best-fit power laws match quite closely
the PD nonlinear corrections for the density $P(k)$.
This match should not be taken as a very meaningful estimate of the
actual shape of the power spectrum in the nonlinear regime.
Our nonlinear segment results from applying a procedure that is still 
based on the linear gravitational instability relation between velocity 
and density to a phenomenon that is fully nonlinear.
We have no immediate reason to expect the PD formalism to extend to peculiar 
velocities so straightforwardly.  We therefore take this match only as a 
crude guideline to encourage further investigations into the
behaviour of the velocity power spectrum (and the likelihood function) in the
mildly nonlinear regime.

As expected, the nonlinear correction for the M3 catalog is larger
than for the SFI data, because the former has more galaxies nearby
and therefore a larger number of close pairs with small errors.  
The M3 $P(k)$, which was somewhat higher than SFI in
the linear analysis, becomes somewhat lower than SFI as a result
of the nonlinear analysis, but the two catalogs basically
yield consistent results.
The likelihood improvement for M3 is very significant,
$\Delta \ln \L \simeq 22$, while for SFI it is moderately significant,
$\Delta \ln \L \sim 2.8$.

\fig{likeomega} shows the likelihood as a function of $\omm$ for each
of the two datasets, where for each value of $\omm$, the power-law 
parameters in the nonlinear regime obtain their most likely values.
The maximum is narrower for M3 than for SFI.
When, instead, we marginalize over the two power-law parameters 
in the nonlinear regime, the obtained best $\omm$ remains the same (to within
$0.01$) and the likelihood function becomes wider by $4\%$.

Although we may expect the position of the break in the power spectrum, 
$\kb$, to be in the vicinity of $k\sim 0.2$ (\eg, from the PD approximation),
we should check the robustness of our results to the actual choice of $\kb$.
\fig{omega_by_k} shows the derived values of $\omm$ and the corresponding
likelihood as a function of the value of $\kb$.
We find that the results are quite insensitive to the choice of $\kb$
over a wide range.
At $\kb$ values much smaller than $0.1$, 
corresponding to large separations between pairs of objects and thus involving
mostly distant objects of large errors, there are
insufficient data to constrain the power spectrum,
and therefore the errors become big and the results quite meaningless.
At very large values of $\kb$, the analysis is expected to recover the
results of the old linear analysis with no break. It indeed does so, but
only when $\kb$ approaches the artificial cutoff
applied to $P(k)$ arbitrarily at $k_{\rm max}=8\ihmpc$ 
for the purpose of finite numerical integration.
It seems that any little freedom allowed in the model beyond the strict linear 
power spectrum is enough for correcting the bias associated with the linear 
analysis.

As mentioned earlier, an alternative way to incorporate nonlinear effects 
is by adding to the linear velocity correlation model a free parameter of 
uncorrelated velocity dispersion at zero lag, $\sigv$. 
When this nonlinear correction is applied by itself
to the M3 data, the best value of $\omm$ becomes $0.38$ (instead of 0.56
in the linear analysis) with $\sigv=250\kms$.

\begin{figure*}[t!]
\vspace{8.5truecm}
\includegraphics{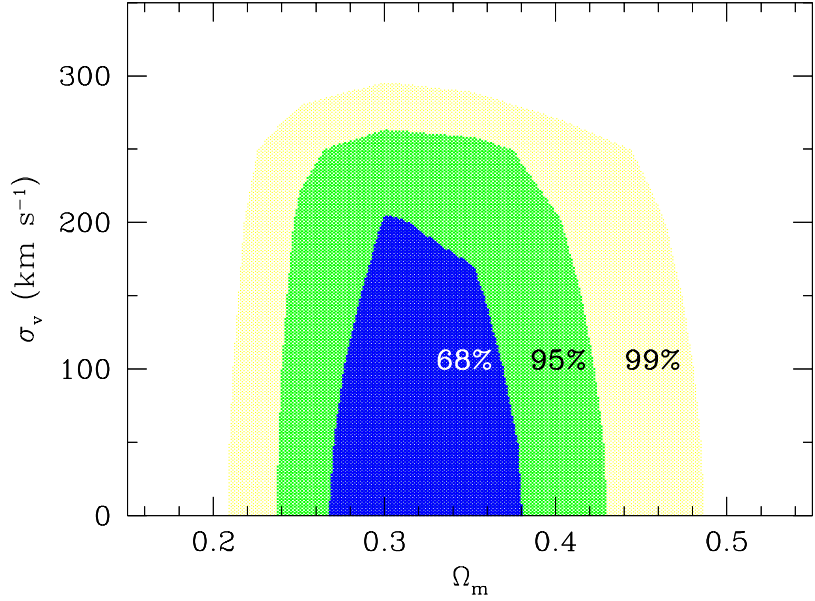}
\caption{\protect \capt
Robustness to different nonlinear corrections.
The likelihood, for the real M3 data, when allowing both a break at
$k=0.2$
and a velocity-dispersion term, as a function of the free parameters
$\omm$ and $\sigv$.
The contours correspond to 68\%, 95.4\% and 99.73\% in the two-parameter
plane.
}
\label{fig:break+sigv}
\end{figure*}

We then apply to M3 the two different nonlinear corrections together,
i.e., a break in the power spectrum at $k=0.2$ as well as a free velocity
dispersion term.  \fig{break+sigv} shows a map of the resulting likelihood 
in the $\omm$-$\sigv$ parameter plane. The best-fit value of $\sigv$ is 
close to zero, indicating that the two nonlinear corrections are practically 
redundant.

\subsection{Deviations from \lcdm}
\label{sec:results_bins}

Encouraged by the success of breaking the power spectrum into two
detached segments, we now push the idea further, and divide the power
spectrum into 4 detached segments.  This allows a more general shape
for $P(k)$, less dependent of \apriori\ assumptions about a physical
model such as \lcdm.  By doing so we may detect clues for deviations
from the ``standard" $P(k)$ shape, but in this case we clearly give up
the attempt to determine cosmological parameters.

Our 4-band model for $P(k)$ consists of the following segments: 
\begin{enumerate}
\item COBE-normalized \lcdm\ in the extreme linear regime, 
$k \leq 0.02$, with $\omm$ fixed at the most likely value from the
nonlinear analysis. 
\item A free constant amplitude in the interval $0.02<k\leq0.07$, 
at the vicinity of $k_{\rm peak}$. 
\item An independent free constant amplitude in the interval $0.07<k\leq0.2$,
just short of the transition between the linear and nonlinear regimes.
\item A power law with two free parameters (as before) in the nonlinear
regime, $k>0.2$.
\end{enumerate}

\begin{figure*}[t!]
\vspace{10.0truecm}
\includegraphics{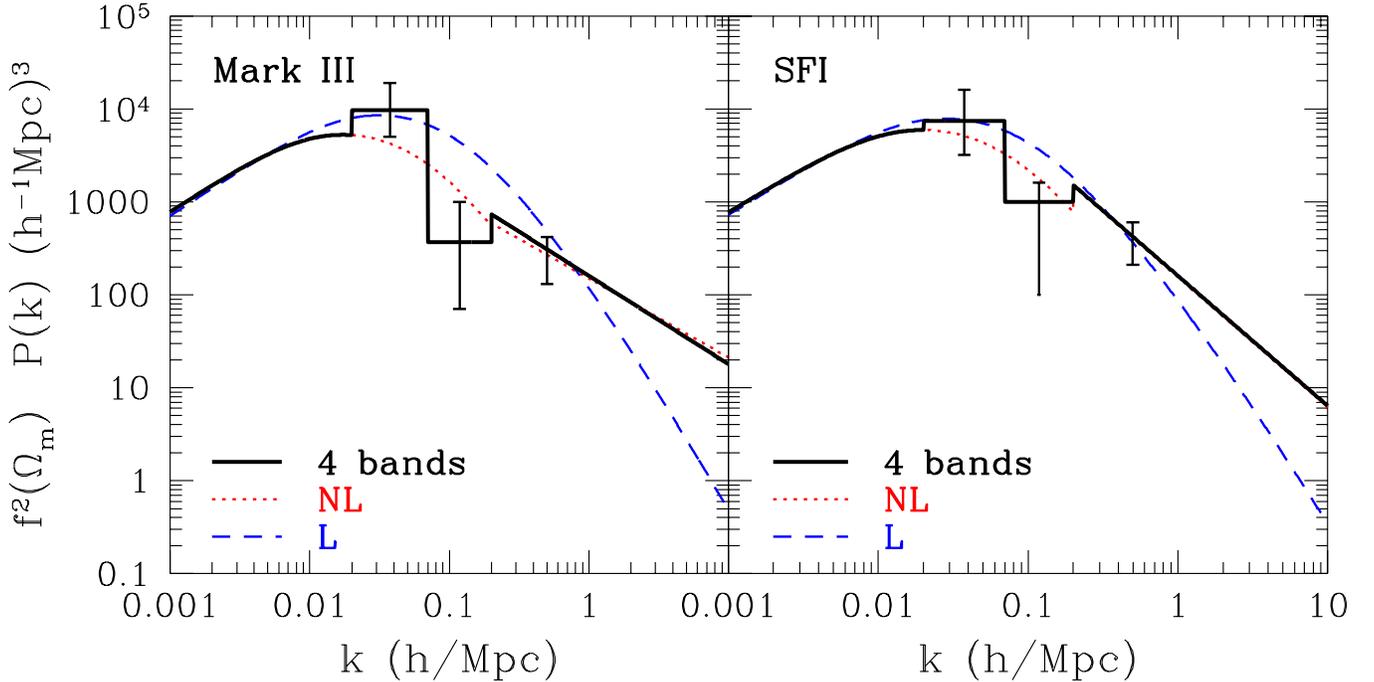}
\caption{\protect \capt
The 4-band power spectrum for the M3 (left) and SFI (right) data,
compared to the best-fit \lcdm\ power spectra, linear (L) and nonlinear
with a break (NL).
}
\label{fig:4steps}
\end{figure*}

\fig{4steps} shows the recovered 4-band $P(k)$ from the real data of M3 and 
SFI, in comparison with the \lcdm\ results of the linear and nonlinear analysis
discussed above.
The most likely parameters of the three meaningful segments ($k>0.02$) are 
$(9750,\, 370,\, 160k^{-0.95})$ for M3 and   
$(7450,\, 1000,\, 160k^{-1.40})$ for SFI,    
where the amplitudes are in unites of $\!\3hmpc$ and $k$ is in units of
$\!\ihmpc$.  The errors are shown in the figure. 

The nonlinear segment, not surprisingly, practically recovers the results 
of the broken-\lcdm\ analysis. 
The most linear segment can be fixed almost arbitrarily without affecting
the results in the two inner bands.
This is not surprising either since we cannot expect our current data to 
contstrain the power spectrum on these very large scales.
 
The second segment in the linear regime, $(0.02,\,0.07)$, 
tends to lie above the peak obtained in the broken-\lcdm\ analysis,
while the third linear interval, $(0.07,\,0.2)$,
in the ``blue" side of the peak, shows a low amplitude.
We thus see a ``wiggle" about the linear portion of the broken-\lcdm\ best fit,
which is detected independently in the two data sets.
The features in the linear regime contribute only a marginal improvement 
to the overall likelihood, which is still dominated by the nonlinear segment. 
The significances of the deviations, both the excessive peak and 
the low dip, are 
slightly above 1$\sigma$ each. 
Our finding should therefore be considered as a marginal hint only;
it could be just a fluke due to the distance errors and cosmic variance.
But it is still an intriguing feature, especially since it appears 
consistently in our two samples. 
%
 
The marginal deviation from the broken-\lcdm\ $P(k)$ thus consists
of a wiggle, with a power excess near the peak, $k\sim 0.05$,
and a deficiency at $k \sim 0.1$.
The missing power is reminiscent of the indications for ``cold flow"
in the galaxy peculiar velocity field in the local cosmological
neighborhood. 
While the streaming motions 
on scales of a few tens of megaparsecs are on the order of several
hundreds of kilometers per second, the dispersion velocity of field
galaxies is only on the order of $\sim 200\kms$, indicating a high
Mach number on comparable scales
(\eg, Suto, Cen \& Ostriker 1992; Chiu, Ostriker \& Strauss 1998; 
Dekel 2000). 

A hint for a similar wiggly feature has been detected in the density
$P(k)$ as derived by some of the researchers from the distribution 
of galaxies (Baugh \& Gazta\~naga 1998; Landy \etal 1996) and clusters 
(Einasto \etal 1997; Suhhonenko \& Gramann 1999). 
Most recently, there are indications for such a wiggle in the
preliminary $P(k)$ derived from part of the 2dF redshift survey (private 
communication with the 2dF team). 

Most interestingly, the scale of the missing power in our local $P(k)$
from velocities roughly coincides with
the scale of the second peak in the angular spectrum of the CMB.
Preliminary balloon measurements 
(Boomerang, de Bernardis \etal 2000; Maxima, Hanany \etal 2000)
indicate that 
this peak is somewhat lower than expected by the common CDM models. 
This may be a reflection of the same phenomena which we detect here as 
``cold flow" in the peculiar velocity data.

The scale of the wiggly feature roughly coincides with the most obvious
physical scale in cosmology --- the size of the cosmological horizon
at the time of transition from radiation to matter dominance, or
slightly later, at the epoch of plasma recombination 
and radiation-matter decoupling.  A wiggle on these scales
can be produced by an excess of either baryons or massive neutrinos
in the cosmological mass budget. But the excess required to
produce a significant wiggle seems to violate upper limits from
other data; the density of baryons is limited by He+D abundances
via the theory of Big-Bang nucleosynthesis (Tytler \etal 2000), 
and the density of neutrinos is constrained by large-scale structure 
(\eg, Ma 1999; Gawiser 2000).

\section{PRINCIPAL COMPONENT ANALYSIS}
\label{sec:pca}


The linear \lcdm\ analysis of both the M3 and SFI data (Z97; F99) revealed 
a warning signal concerning the GOF, which we termed ``the two-halves problem".
When the linear analysis is applied separately to two halves of the data, 
separated either by the median distance or by line-width (which is correlated
with the distance), the results are somewhat different.
The distant data prefer a lower-amplitude power spectrum, associated with a 
lower value of $\omm$. 
The mock catalogs (tested with the true correlation
matrix) have not revealed a similar problem, indicating that it is
caused by inadequacies of the correlation matrix compared with the real data.
These shortcomings may be associated with the assumed theoretical
model, either the shape of the $P(k)$ or the Gaussian nature
of the fluctuations, or with the error model,
either its Gaussianity or its radial dependence.
These worries motivate an attempt to evaluate the GOF in our linear 
and nonlinear analyses.
In particular, we wish to see to what extent the revised $P(k)$ in the 
nonlinear regime may resolve the two-halves problem.

Assume a data vector $\bd$, which is a random realization of an
$n$-dimensional multivariate Gaussian distribution, with the correlation
matrix $C=\langle \bd \bd^{\dagger }\rangle$.
A global GOF could be evaluated using the $\chi^2$ statistic,
$\chi^{2} = \bd ^{\dagger} C^{-1} \bd = Tr(C^{-1}D)$,
where $D\equiv \bd \bd^{\dagger}$.
If $C$ is the true correlation matrix, then this quantity should
obey a $\chi^2$ distribution with $n$ degrees of freedom,
as indeed is the case for all the analyses we performed.
But this
single number cannot capture all the particulars of the fitting process.

A Principal-Component Analysis, in which the data are represented
in terms of the eigenvector basis of the (assumed) correlation matrix,
is a powerful tool for our purposes in several different ways. 
Our original motivation for applying a PCA (following Vogeley \& Szalay 1996;
Tegmark, Taylor \& Heavens 1997) 
was to allow optimal compression of the data into the modes 
that are most important for estimating the parameters we wish to evaluate, 
with the aim to reduce the computational cost associated with inverting 
huge correlation matrices, and to improve the results given an inaccurate 
correlation matrix.  
In the current paper, we use a PCA for two other purposes.
First, we present a novel approach of using the PCA modes 
for identifying certain gross features of the data and model via the
correlation matrix. Second, we extend a method first used by
Hoffman \& Zaroubi (2000) for evaluating GOF in fine detail\footnote{
In the terminology which is introduced later, they used the particular 
case of cumulative $\chi^2$ per degree of freedom for $S+N$ modes, and we
extend the analysis to differential $\chi^2$ and to $S/N$ modes.}
and for trying to resolve the two-halves problem.
%

\subsection{Modes of $S+N$ versus $S/N$}

\begin{figure*}[t!]
\vspace{11.0truecm}
\includegraphics{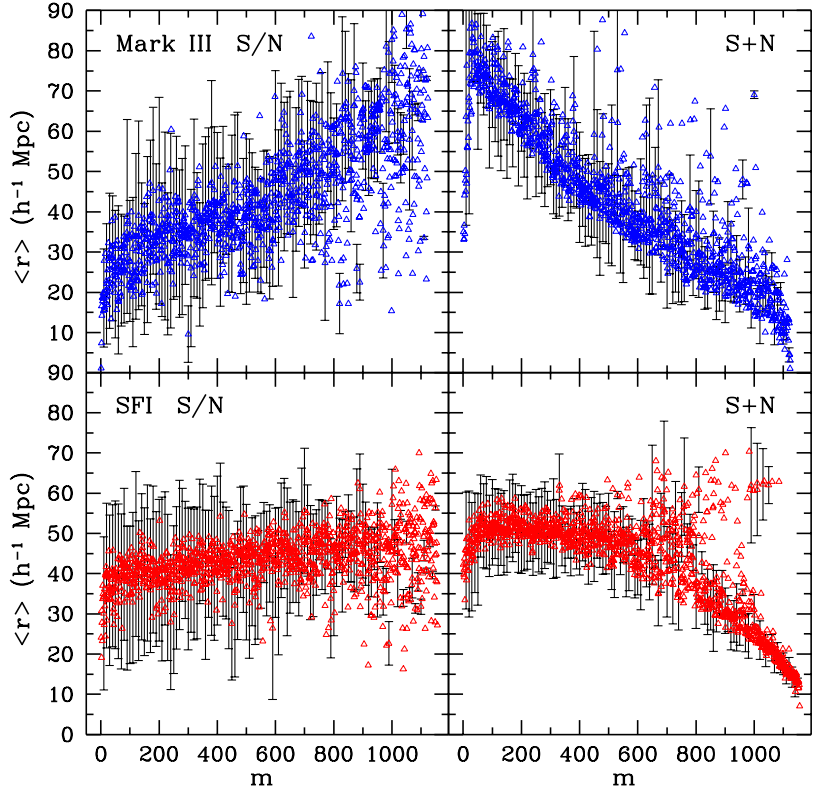}
\caption{\protect \capt
Average distances associated with the eigenmodes of the correlation
matrix
of the linear \lcdm\ model.
The eigenmodes are ranked by decreasing eigenvalue amplitude.
(low $m$ --- high eigenvalue).
The standard deviation is shown for every tenth mode.
Left: $S/N$ modes. Right: $S+N$ modes.
Top: M3 data. Bottom: SFI data.
}
\label{fig:r}
\end{figure*}

The standard PCA is as follows.
A general coordinate transfromation of the data $\bd$
defines a new set of $m$ random variables
$\tilde{\bd}=\Psi \bd $, where $\Psi$ is an $m\times n$ matrix, 
which we assume to be of full rank. It is then clear that the
distribution of the new variables is still Gaussian, with a correlation
matrix
$\widetilde{C}=\langle \widetilde{\bd}\widetilde{\bd}^{\dagger}
\rangle =\Psi C\Psi ^{\dagger}$. 
The likelihood analysis can be performed in terms of these new ``data" points.
If we keep all the original information, \ie, if $\Psi$ is invertible, 
the likelihood analysis is not affected because then
$Tr(\tilde{C}^{-1}\tilde{D}) =Tr(C^{-1}D)$.
Since the correlation matrix is symmetric and positive definite, we can,
without loss of generality, pick the matrix $\Psi$ such that $\tilde{C}$ 
is diagonal.  Then the rows $\tilde{d}_{i}$ of $\Psi$ are the eigenvectors
of the correlation matrix, or its principal components, 
and the diagonal terms $\lambda_i$ of $\tilde{C}$ are the corresponding
eigenvalues,
In statistical terms this means that the new variables are expected
to be uncorrelated.  The validity of this independence of variables 
is a measure of GOF.  We thus use the $\chi^{2}$ statistic to test the 
hypothesis that $\tilde{d}_{i}/\sqrt{\lambda _{i}}$ are uncorrelated unit 
Gaussian random variables.\footnote{
Another advantage of having the modes uncorrelated
is that, when compressing the data, it makes sense to have no
correlation between the data kept and the data eliminated.}
If this test uncovers systematic effects, it may become possible to
associate them with certain features of the data and model 
via a further investigation of the eigenmodes. 

The eigenmodes are ordered by the amplitude of their eigenvalues, 
from large to small, and the high-eigenvalue modes are assigned a higher
signifcance, because the confidence levels in the recovered 
parameters of a maximum-likelihood analysis inversly correlate with 
the squares of the eigenvalues of the modes used (Tegmark,
Taylor \& Heavens 1997). 
Furtheremore, perturbation analysis implies that small-eigenvalue modes 
are more sensitive to perturbations in the correlation matrix, implying that
the mode-by-mode statistical tests may not be reliable 
for small-eigenvalue modes.  Since our correlation matrix is expected to
be only an approximation to the true correlation matrix, it would be 
advantageous, in general, to avoid small-eigenvalue modes, and rely
on the high-eigenvalue ones.

A straightforward application of PCA is with the original correlation
matrix of \equ{C}, which is a sum of signal and noise:
$C=S+N$. In this case, the large eigenvalues may correspond
either to large signal, or large noise, or both.
Another possibility, which we term $S/N$,
is to first perform a ``whitening" transformation,
$\hat{\bd}=N^{-\frac{1}{2}}\bd$ (Vogeley \& Szalay 1996).
In the case of a diagonal noise matrix $N$, this transformation
amounts to normalizing the data in terms of the expected noise.
The new correlation matrix is
$\hat{C}=N^{-\frac{1}{2}}SN^{-\frac{1}{2}}+I$, where $I$ is
the identity matrix.
The eigenvalues of $\hat{C}$ are the signal-to-noise ratios of the
corresponding principal modes.\footnote{
It seems, therefore, that the $S/N$ modes can provide a proper basis for
optimal data compression.  
As said before, if the models for the signal and errors are perfectly accurate, 
then the truncation would not affect the result while
the estimated uncertainties will grow.
However, if the correlation matrix is only approximate,
using the high-$S/N$ part of the data may improve the results. 
In particular, since the correlation matrix is quadratic in the data, 
an error in the error model would necessarily lead to a systematic error 
in the results. By eliminating the low-$S/N$ modes 
such systematics may be reduced.}

\subsection{Correlation Between Mode and Distance}

The eigenmodes can help us identify certain features of the data and models.
In particular, the signal part involves the geometry of sampling and
the prior model, and the noise part involves the distance error estimates.
A useful diagnostic statistic to assign with each eigenmode is the distance 
from the Local Group. For eigenvector $v$, the average distance is 
\be
\langle r\rangle _{v}=\sum_g |v(g)|^{2} r(g)\, ,
\label{average_r}
\ee 
where the sum is over the sample of galaxies, $r(g)$ is the distance 
of galaxy $g$, and $v(g)$ defines the vector $v$ in the basis $g$. 
This is a weighted average of the galaxy distances. 
The variance $\left\langle (r- \langle r \rangle_v)^2 \right\rangle_v$
is defined in analogy.
If the standard deviation is small compared to the average distance, 
one can conclude that most of the information associated with this mode 
comes from galaxies within a certain distance range.
 
The distance associated with a mode provides important information
about the mode: distant modes are typically noisier than nearby ones
because the distance error is proportional to distance. 
The correlation between modes and distance could also 
help us understand the two-halves problem.
In the following mode-by-mode analysis, we use this correlation to interpret
a correlation with mode number as a correlation with distance.

\fig{r} shows the average distance for each mode, for the linear \lcdm\ 
model and either the M3 or SFI data.
We see that the $S/N$ modes are correlated with distance,
such that the high-eigenvalue modes, which are robust and of high
signal-to-noise ratio, are typically associated 
with nearby data, which are of relatively small errors.
On the other hand, these nearby modes tend to involve close galaxy pairs,
and are therefore more subject to nonlinear effects, which makes the
nonlinear correction a must.
The correlation is strong for M3, and weaker for SFI.

The $S+N$ modes show a somewhat stronger correlation
in the opposite sense, in which the high-eigenvalue modes, except for
the first few, are typically associated with large distances and therefore
noisy data. This means that most of the $S+N$ modes are dominated by the noise
rather than the signal.  This situation is unfortunate for the $S+N$ PCA;
for example, it does not allow a sensible truncation by $S+N$ modes.
But it should allow a more sensitive measure of GOF, refering in
particular to the error model. Again, the 
correlation is stronger for M3 than for SFI.

\subsection{Goodness of Fit Mode by Mode}

After PCA, assuming that we know the true correlation matrix, 
the variables $\widetilde{\bd}$ are expected to be uncorrelated,
and we expect $\chi ^2_i = \tilde{d}_i^2/\lambda _i$
to be about unity for each and every mode separately. 
The validity of this behavior mode by mode provides an improved
and finer test of GOF in two ways.
First, it tests whether the eigenmodes of the prior correlation matrix 
are really uncorrelated, with the variance determined by the eigenvalues.
Then, in the case of a poor fit for a certain mode, it can guide us to the
source of the poor fit via the properties associated with that mode.

One statistic we use, for each mode number $m$, is the cumulative 
$\chi^2$ per degree of freedom, $\sum ^m_{i=1}\chi _i^2/m$, in which
the sum starts from the high-eigenvalue modes and ends at mode $m$. 
In the case of independet modes,
the expected value is unity, and the expected standard deviation is
about $\sqrt{2/m}$ (the normalized standard deviation of a $\chi ^2$
distribution with m degrees of freedom).

A second set of more localized statistics is the differential estimates,
obtained by averaging the $\chi^2$ values over intervals of
mode numbers of length $s$ namely, using $\sum_{i=m-s+1}^{m}\chi_{i}^{2}/s$
for various values of $m$.
If the correlation matrix was exact, these would be independent
(assuming the intervals are disjoint) and follow a $\chi^2$ 
distribution with $s$
degrees of freedom. In particular, the expectation value 
would be unity with a standard deviation $\simeq \sqrt{2/s}$.
We choose $s=50$, which is large enough for good statistics and
small enough with respect to the total $n$ for the purpose
of tracing systematic effects.

\begin{figure*}[t!]
\vspace{10.0truecm}
\includegraphics{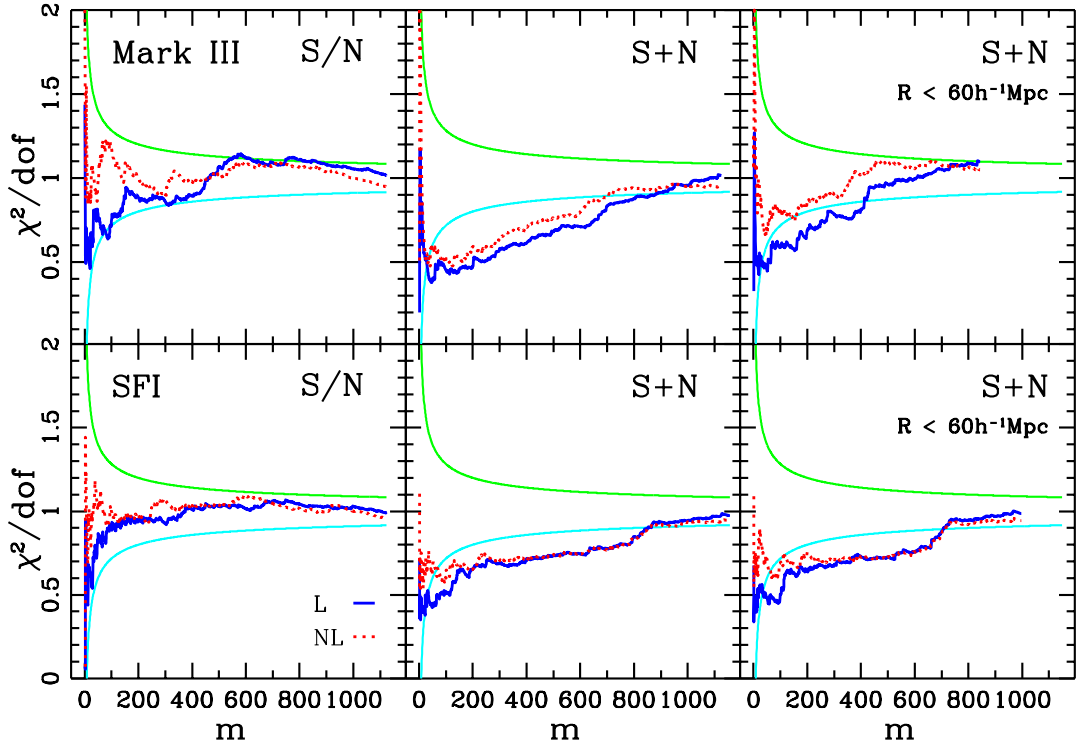}
\caption{\protect \capt
Cummulative $\chi ^2$ per degree of freedom as a function of mode
number. The
heavy
solid curve is for the linear \lcdm\ model, and the dotted
curve is for the broken-\lcdm\ power spectrum.
The two solid lines mark the expected $2\sigma$ deviations.
Left: $S/N$ modes.  Middle: $S+N$ modes.
Right: $S+N$ modes, for the data at $r< 60\hmpc$ only.
Top: M3 data. Bottom: SFI data.
}
\label{fig:chi2_cum}
\end{figure*}

\begin{figure*}
\vspace{10.0truecm}
\includegraphics{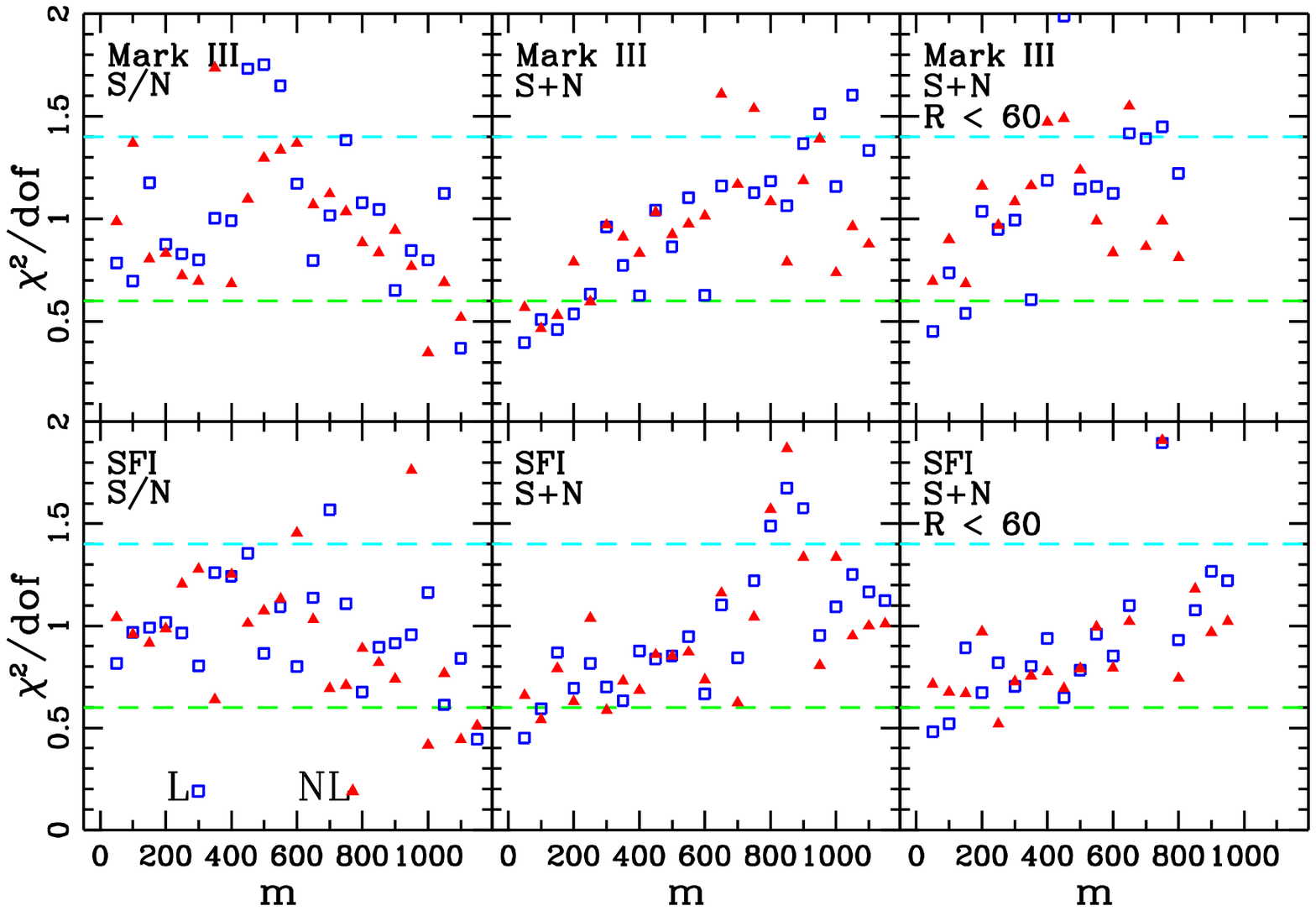}
\caption{\protect \capt 
Differential $\chi ^2$ per degree of freedom as a function of mode number.
The open squares are for the linear \lcdm\ model, and the filled triangles 
are for the broken-\lcdm\ power spectrum.
The two dashed lines mark the expected $2\sigma$ deviations.
Left: $S/N$ modes.  Middle: $S+N$ modes.
Right: $S+N$ modes, for the data at $r< 60\hmpc$ only.
Top: M3 data. Bottom: SFI data.
}
\label{fig:chi2_dif}
\end{figure*}

\fig{chi2_cum} shows the cumulative $\chi^2$ statistic as a function of $m$
for the linear \lcdm\ analysis 
and for the nonlinear broken-\lcdm\ analysis.
\fig{chi2_dif} shows the corresponding differential $\chi^2$ statistic.

For the $S/N$ modes of M3, the GOF of the linear model is marginal,
in the sense that the typical deviations in the cumulative statistic
are at the $2\sigma$ level.
The low-$m$ modes, except for the very first ones, typically
have low $\chi^2/dof$ values, while the large-$m$ modes have high values.
This systematic behavior with $m$ can be translated to a systematic
correlation with distance via the correlation between distance and mode
(\fig{r}). It is therefore a reminiscence of the two-halves problem. 

Can we use the PCA to distinguish between an inadequate model of $P(k)$ 
and a problem in the error model?
We see in \fig{chi2_cum} that
the broken-\lcdm\ model clearly improves the GOF as far as the
$S/N$ modes of M3 are concerned. With this model, the cumulative $\chi^2/dof$
lies well inside the $2\sigma$ contours for all the modes, with no
apparent systematic dependence on $m$.  It implies that the broken-\lcdm\
$P(k)$ is a more appropriate model for the data. Based on the $S/N$ modes, 
there is no indication that the error model may be inadequate.

When we analyze the $S+N$ modes in a similar way in \fig{chi2_cum}, 
the linear model, 
for both data sets, 
shows a more severe deviation of $\chi^2/dof$ from unity, at the 
$4-5\sigma$ level, and a similar systematic dependence on $m$. 
This trend is also apparent in \fig{chi2_dif} (middle panels). 
The two-halves problem is very obvious here, with the more
distant data, corresponding to larger eigenvalues and larger noise,
favoring a smaller amplitude for the power spectrum than the nearby data. 
In this case, the use of the better, broken-\lcdm\ model makes only a 
small improvement which does not resolve the problem.  
This is a clear indication that something may be wrong in the error model 
as well.

We then recall that the low-$m$ $S+N$ modes are associated with large 
distances, where the errors are large and are known to a lesser accuracy.
Guided by \fig{r}, we try a poor-man compression of the data
by eliminating from the analysis
all the data points that lie at an inferred distance greater than $60\hmpc$.
This leaves us with $843$ out of the $1124$ (grouped) data points of M3, 
and $996$ out of the $1156$ galaxies of SFI.  
This truncation makes only a negligible change in the 
the best-fit value of $\omm$ (an increase of less than $3\%$, both for M3 
and SFI), and it causes only a minor widening of the likelihood contours.
In the case of M3, we see in \fig{chi2_cum} that
the $S+N$ modes of the linear model and truncated data
show an improved GOF compared to the case of the whole data,
but the $\chi^2/dof$ still show $\sim 3\sigma$ deviations from unity
and a systematic dependence on $m$.
However, the $S+N$ modes of the broken-\lcdm\ model and truncated M3 data 
now do  lie within the $2\sigma$ contours.  The systematic trend with $m$ is
still apparent, indicating that the correlation matrix is still not perfect;
either the error model is only an approximation even for the 
truncated data, or the broken-\lcdm\ $P(k)$ is not
yet a perfect model (as seen in \se{results_bins}), or
the signal and/or the noise are not exactly Gaussian.
 
In the case of SFI, 
while the $S/N$ modes look very adequate with both models,
for the $S+N$ cumulative statistic the improvements due to the nonlinear 
correction and the elimination of large-distance galaxies are apparently not 
enough for an acceptable GOF.  
According to the differential statistic, the
nonlinear correction and truncation do bring each of the first few bins
into their $2\sigma$ range, but the fact that many of 
these bins are each not much above the $-2\sigma$ line makes the cumulative
statistic lie outside its $2\sigma$ range.
Since the large-eigenvalue $S+N$ modes, which dominate the cumulative
statistic, are dominated by noise, the limited GOF is likely to point
at further shortcomings of the error model for SFI.

\section{CONCLUSION}
\label{sec:conc}

A likelihood analysis is supposed to recover unbiased values for the
free parameters of a model provided that the prior theoretical model 
and the error model allow accurate description 
of the data. We addressed here tools to recover the parameters given 
incomplete knowledge of these models.

Using mock catalogs based on high-resolution simulations, we realized
that the likelihood analysis of peculiar-velocity data, based on the 
linear \lcdm\ power spectrum, is driven by the nonlinear part of the 
spectrum which is not modeled accurately, and might 
therefore yield biased results. For example, in the linear analysis of 
the M3 mock data, the obtained amplitude of $P(k)\omm^{1.2}$ is overestimated, 
corresponding to a positive bias in the cosmological density parameter $\omm$ 
by $\sim 35\%$.

A broken-\lcdm\ power spectrum, in which the $k>\kb$ segment of the power 
spectrum is replaced by a more flexible two-parameter power law, 
allows a better, independent fit in the nonlinear regime. It then
frees the linear part of the spectrum from nonlinear effects, and
yields unbiased results for $\omm$. The results are robust to the 
specific choice of $\kb$; we choose $\kb = 0.2\ihmpc$, which is
where the nonlinear density $P(k)$ is expected to start deviating from 
the linear $P(k)$ by the PD approximation.
The results are also robust to the exact way by which the nonlinear
effects are incorporated. When we add a zero-lag velocity dispersion 
term to the correlation function, either replacing the break in the
power spectrum or in addition to it, the results are similar.

We note that these procedures do not mean to tell us much about the
actual power spectrum in the nonlinear regime, as our improved analysis
in the nonlinear regime is still based on the linear relation betweem 
velocity and density. 
For an improvement in this direction, one should come up with a physically
motivated functional form that accounts for mildly nonlinear corrections
to the linear velocity power spectrum, in analogy to the PD correction
to the density $P(k)$.
Nevertheless, as demonstrated by the tests based on mock catalogs
and by the robustness to the actual nonlinear correction applied,
our current procedures are reliable for eliminating the biases
in the linear regime.  

When applied to the real data of M3 or SFI peculiar velocities,
assuming 
a flat \lcdm\ model with $n=1$ and $h=0.65$, 
the improved analysis yields best-fits of 
$\omm = 0.32\pm 0.06$ and $0.37\pm 0.09$ respectively,
corresponding to 
$\sigma _8\omm ^{0.6}\approx 0.49 \pm 0.06$ and $0.63 \pm 0.08$ 
respectively. These values are in good agreement with most constraints 
from other data, including CMB anisotropies and cluster abundance 
(\eg, Bahcall \etal 1999).\footnote{ 
Joint analyses of peculiar velocities with other dynamical data free
of galaxy biasing were pursued before based on the linear analysis 
(Zehavi \& Dekel 1999), and now based on the nonlinear analysis 
(Bridle \etal 2001).}
It is important to stress that the estimates of the cosmological
parameters reported here are valid under the assumption that the 
power spectrum in the liner regime is drawn from the flat 
\lcdm\ cosmolgical model with $\omm$ as a free parameter.
In our current nonlinear analysis of the data inside $60\hmpc$,
the maximum-likelihood solution based on the \lcdm\ power spectrum
indeed has an acceptable goodness-of-fit, making our estimate of
$\omm$ self-consistent and meaningful. 
Although we conclude below that our error model is not accurate
for M3 beyond $60\hmpc$ and for SFI, the obtained value of $\omm$ is
insensitive to the exclusion of the noisy data beyond $60\hmpc$ and
to other changes in the error model.

By allowing 
a 
more general shape for the power spectrum, with 4
detached segments, we detect a
marginal indication 
for a deviation from the \lcdm\ power spectrum. 
The possible deviation 
is characterized by a
wiggle, with an enhanced amplitude near $k_{\rm peak} \sim 0.05$
and a depletion near $k \sim 0.1\ihmpc$.
The study of possible deviations from the \lcdm\ model is done
at the expense of the attempt to estimate $\omm$, which requires
a power-spectrum shape based on a physical cosmological model.
Nevertheless, we learn from the 4-band analysis that the relatively low 
$\omm$ estimate in the broken-\lcdm\ analysis is driven by the low
amplitude of the power spectrum near $k \sim 0.1\ihmpc$.

The 
indicated 
``cold flow" on a scale of a few tens of megaparsecs
is reminiscent of similar indications from the power spectrum
of galaxies and clusters in redshift surveys (\se{results_bins}).
Most recent is the wiggle seen in the preliminary power spectrum derived  
from the 2dF redshift survey.  The local cold flow may be related to the 
second peak in the CMB angular power spectrum on a similar scale.
The wiggly feature in the power spectrum may be interpreted
as a possible indication of a deviation from the standard cosmological
mass mixture, \eg, a higher baryonic content than indicated by
the Deuterium abundance and Big-Bang nucleosynthesis, or a non-negligible
contribution from hot dark-matter in the form of massive neutrinos.
However, the possibility that this feature is a statistical 
fluke due to cosmic
variance in the context of the \lcdm\ model cannot be ruled out yet.

A principal component analysis, either in $S/N$ or $S+N$ modes, allows a 
fine test of goodness of fit, by applying a $\chi^2$ test mode by mode.
It shows that the broken-\lcdm\ model is a better fit to the data than
the purely linear \lcdm\ model.  For M3, using the ``whitened" $S/N$ modes, 
the nonlinear correction is enough to eliminate the ``two-halves" problem
that troubled the linear analysis. 
This indicates that the \lcdm\ model is a good model for the signal.
When the $S+N$ modes are analyzed,
the correction to the theoretical model is helpful but 
not enough for an acceptable GOF.
When the M3 data is further truncated at $60\hmpc$, eliminating 
distant galaxies for which the errors are large and the error model is 
inaccurate, the GOF becomes acceptable. 
This strengthens the reliability of our parameter estimates based on
the assumed \lcdm\ model in the linear regime.
Future investigations should refine the error model at large distances,
in order to possibly reduce the cosmic variance in our estimates.
For SFI, the $S/N$ modes seem adequate, 
confirming again the suitability of the \lcdm\ model in the linear regime, 
but the $S+N$ PCA indicates that the errors 
in this catalog 
are still more complex than assumed.

The PCA is a powerful tool for addressing interesting properties
of the data and its relation to the best-fit theoretical and  
error models. In particular, we associated each mode with a geometric
property --- the mean distance and the variance about it --- 
and thus learned about the correlation between mode eigenvalues 
and distance errors. This was useful in the study of GOF and in
truncating the data to deal with inaccuracies in the error model.
The PCA will be extremely useful when one tries to compress the data
while keeping the optimal part for determining a specific desired parameter.
This compression may be mandatory for computational reasons
when the body of data is excessively large.
Since the model is expected to be incomplete,
either in terms of the theoretical assumptions or the errors,
a proper compression of the data may in fact improve the results.
Such data compression using PCA in the context
of the analysis of cosmic flows is in progress.

Our conclusions can be summarized at three levels, as follows.
First, if one is willing to {\it assume\,} the ``standard" \lcdm\ 
power-spectrum shape in the linear regime, then the nonlinear corrections yield 
an unbiased value of $\omm \sim 0.35$, consistently from the two data sets.
The GOF is acceptable for this nonlinear analysis and the cosmological
model and errors assumed, once the M3 data inside $60\hmpc$ 
are considered (and the value of $\omm$ is insensitive to the inclusion 
of the other, noisy data).
This makes the parameter estimation self-consistent and meaningful,
and this is our strongest result.
Second, once we alleviate the requirement of a physically motivated 
power spectrum and its dependence on cosmological parameters,
we detect a marginal indication for a wiggle about the \lcdm\ power 
spectrum in the linear regime. This is a marginal detection, which does
not ivalidate the suitability of the \lcdm\ model for the cosmological
parameter estimation performed above. The hint for a wiggle is still
intriguing because it repeats in the two data sets, it seems
to coincide with similar clues from other data, and it may have 
interesting theoretical implications on the nature of dark matter
and the initial fluctuations.
Third, we learn from our newly developed PCA analysis that in order to 
possibly impprove the statistics one should refine the error model for 
the most distant galaxies in M3, and for a larger fraction of the galaxies 
in SFI. But our parameter estimates and the marginal detection of a wiggle 
are driven by the part of the data for which the error model yields an 
acceptable goodness of fit.
%

\section*{Acknowledgments}
This research has been partly supported by the Israel Science Foundation
grant 546/98, 
by the US-Israel Binational Science Foundation grant 98-00217,
and by the DOE and the NASA grant NAG 5-7092 at Fermilab. 
We acknowledge stimulating discussions with Lloyd Knox, Amos Yahil, and 
Saleem Zaroubi.


\end{document}